\documentclass[a4paper]{jpconf}

\PassOptionsToPackage{log-declarations=false}{xparse} %
\RequirePackage[l2tabu,orthodox]{nag}
\usepackage{graphicx}
\usepackage[sort&compress,square,numbers]{natbib}
\RequirePackage{amsmath}
\RequirePackage{mathtools}
\RequirePackage{xfrac}
\RequirePackage{physics}
\RequirePackage[mode=math, range-units=single, separate-uncertainty=true]{siunitx}

\PassOptionsToPackage{hyphens}{url}
\usepackage{hyperref}
\RequirePackage{microtype} %
\usepackage[noabbrev]{cleveref} %
\usepackage{scalerel}
\usepackage{pgf}

\newcommand{\tu}[1]{\ensuremath{_{\mathrm{#1}}}}
\newcommand{\umux}{$\mu$Mux}

\newcommand{\kalpha}[1][\alpha]{K\ensuremath{_{\scaleobj{1.2}{#1}}}}
\newcommand{\lalpha}[1][\alpha]{L\ensuremath{_{\scaleobj{1.2}{#1}}}}

\begin{document}
\title{Hard X-ray Fluorescence measurements with TESs at the Advanced Photon Source}

\author{Tejas Guruswamy$^1$, Lisa M Gades$^1$, Antonino Miceli$^1$, Umeshkumar M Patel$^1$, John T Weizeorick$^1$, Orlando Quaranta$^1$}

\address{$^1$ X-ray Science Division, Advanced Photon Source, Argonne National Laboratory, Lemont IL, USA}

\ead{tguruswamy@anl.gov}

\begin{abstract}
Transition Edge Sensor (TES) spectrometers for hard X-ray beamline science will enable improved X-ray emission and absorption spectroscopy in the information-rich \SIrange{2}{20}{keV} energy range.
We are building a TES-based instrument for the Advanced Photon Source (APS) synchrotron, to be made available to beamline users. 24-pixel prototype arrays have recently been fabricated and tested.
The first spectroscopy measurements using these arrays are promising, with a best single-pixel energy resolution of \SI{11.2}{eV} and saturation energy $>\SI{20}{keV}$.
We present a series of recent X-ray Fluorescence measurements involving transition metal elements and multi-element samples with closely spaced emission lines, in particular a Cu-Ni-Co thin film and a foil of Cu and Hf.
The TES-measured spectra are directly compared to spectra measured with silicon drift detectors at an APS beamline, demonstrating the improved X-ray science made possible by TES spectrometers.
\end{abstract}

\section{Introduction}

Superconducting energy-resolving photon detectors such as Transition Edge Sensors (TESs) provide an order of magnitude better energy resolution than semiconductor-based detectors, while still maintaining the advantages of an energy-dispersive measurement.
In particular, instruments of this type have the unique capability to record time-resolved images with good energy resolution and near-unity quantum efficiency over a wide energy range.
A number of projects are ongoing and have reported success developing and deploying TES arrays for soft X-rays (typically $<\SI{2}{keV}$)~\cite{Doriese2017}, and high-energy X-ray and gamma rays ($>\SI{50}{keV}$)~\cite{Bennett2012,Hatakeyama2014,Mates2017}.
Our ongoing project at the Advanced Photon Source (APS), Argonne National Laboratory is distinct in targeting the \SIrange{2}{20}{keV} hard X-ray energy range, and doing so using entirely lithographically defined absorbers, ensuring scalability and rapid and consistent fabrication.
The complete first-generation instrument, with the cryogenic system, readout electronics, and software, will be made available to beamline scientists and users at the APS.
It will find immediate science applications in X-ray absorption and emission spectroscopy, where the improved energy resolution will allow chemical analysis previously only possible with crystal analysers.
The collection efficiency of the TES array will be of particular benefit to measurements of very dilute and radiation-sensitive samples.
After the upcoming APS Upgrade, we expect the next generations of TES-based instruments can contribute to at least three of the new or upgraded ``flagship'' beamlines, enabling chemically sensitive X-ray microscopy, energy-dispersive diffraction (EDD), and Compton scattering measurements.

X-ray Fluorescence (XRF) is a powerful technique for the identification of the elemental makeup of unknown samples, and for the quantification of their relative abundances.
Hard X-rays ($>\SI{2}{keV}$) are especially useful as they can probe all elements of the periodic table above phosphorus ($Z=15$).
The K-shell photon emission lines of common 3$d$ and 4$d$ transition metal elements like Fe, Ni, and Cu are in the \SIrange{5}{15}{keV} energy range, as are the L-shell lines of the 5$d$ elements.
In samples containing certain combinations of these metals, such as integrated circuits with Hf oxides and Cu~\cite{Choi2011}, or organometallic complexes relevant to biological processes~\cite{Davis2015}, the present semiconductor-based X-ray energy resolving detectors cannot resolve these overlapping lines.
Silicon drift detectors (SDDs) under ideal conditions achieve an energy resolution of \SI{125}{eV}~\cite{Vortex}, for example.
Failing to properly resolve lines can result in an uncertainty in composition determination, or even completely missing the presence of a dilute element.
In partial fluorescence yield X-ray absorption measurements, where the fluorescence yield of a particular emission line is monitored as the incident beam energy is scanned, the inability to measure the flux from one emission line alone limits the usable energy range.
A TES-based spectrometer with an energy resolution of $\SI{10}{eV}$ at \SI{10}{keV} can easily resolve the emission lines of neighbouring elements, allowing consistent identification of all peaks in XRF spectra, and EXAFS (Extended X-ray Absorption Fine Structure) measurements over a very large range.

Our overall instrument design goals are a 100-pixel array, optimized for \SIrange{2}{20}{keV} X-rays and low count rates ($<100$ counts/s per pixel), with $<\SI{10}{eV}$ FWHM resolution.
We have now successfully fabricated and tested 24-pixel prototype devices.
In our lab, we have used these devices to record the emission spectra of samples which have problematic emission line overlaps when measured with an SDD; here, we report on two examples: a Cu-Ni-Co thin film, and a Cu and Hf foil.
We find our prototype TESs with a saturation energy $>\SI{20}{keV}$ can successfully and consistently resolve the previously overlapping emission lines.

\section{Design and Methods}

\subsection{TES}

\begin{figure}[tbh]
  \centering%
  \includegraphics[width=2.9in]{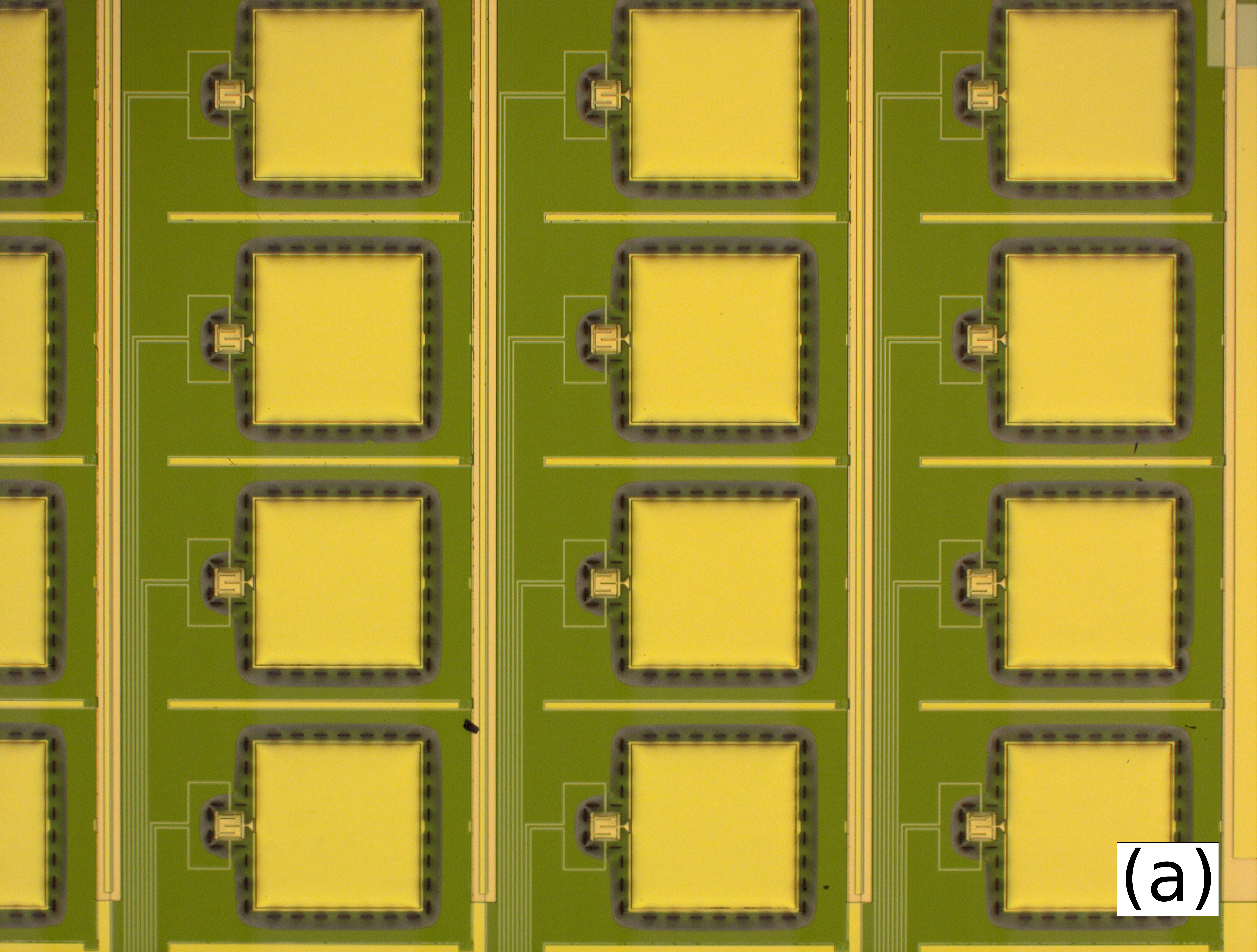}%
  \hspace{0.2in}%
  \includegraphics[width=2.9in]{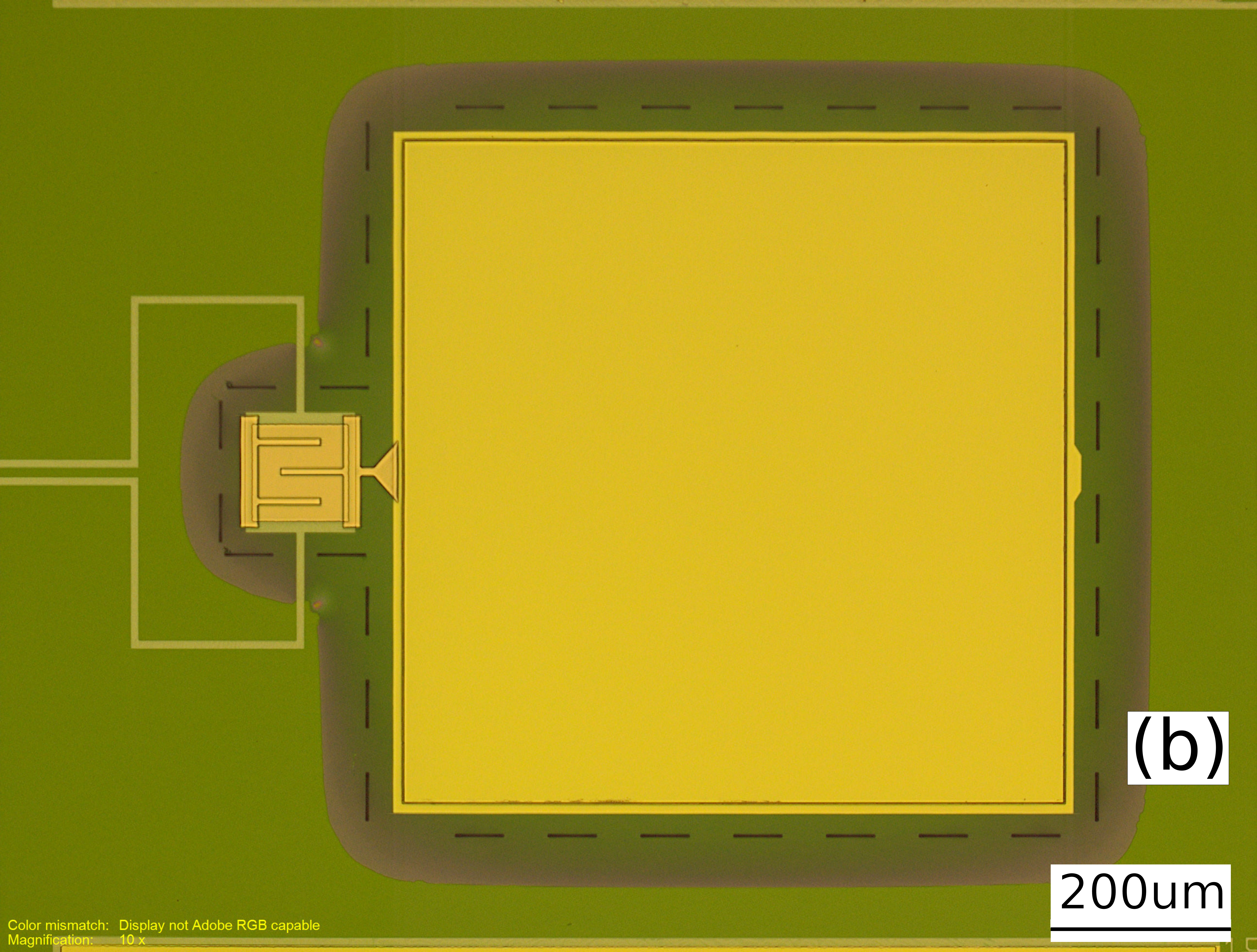}
  \caption{(a)~Optical image of section of 24-pixel TES array. (b)~Optical view of single pixel, showing TES bilayer with Cu features and Au absorber, on perforated SiN$_{x}$ membrane.\label{fig:tes}}
\end{figure}

Each TES pixel is a superconducting bilayer square with a sidecar Au absorber, all on a thin SiN$_{x}$ membrane.
The $(\SI{120}{\micro m})^2$ sputtered Mo-Cu bilayer has additional Cu banks and bars patterned on top to reduce noise~\cite{Ullom2004}, as shown in \cref{fig:tes}.
In the devices reported on here, the absorber is sputtered Au, \SI{1}{\micro m} thick, connected to the bilayer by a Cu patch. %
We estimate this thickness of gold is sufficient to absorb 32\% of incoming X-ray photons at \SI{8}{keV}.
In order to increase the X-ray stopping power of the absorber without significant increases to the heat capacity, the next iteration of devices will be adding electroplated Bi.
Using electroplated rather than evaporated Bi reduces the presence of low-energy tails in measured spectra~\cite{Yan2017a,Yan2018}.
The SiN$_{x}$ membrane is created by back-etching through a Si wafer with Deep-Reactive Ion Etching (DRIE). %
The membrane is also perforated to further reduce the thermal conductance from the TES to the surrounding substrate.
Mo leads ($T_c \sim \SI{1}{K}$) connect the TES to the external circuitry.
Details of our TES fabrication process are available in Ref. \citenum{Patel2019sub}.
Arrays of shunt resistors (\SI{300}{\micro \ohm}) and inductors (\SI{430}{nH}) are on separate chips provided by the Quantum Sensors Group at NIST (National Institute of Standards and Technologies, Boulder, CO, USA), and allow for the simultaneous biasing of all TES pixels with one pair of DC lines.
The stiff voltage bias circuit is completed by a battery-powered isolated voltage source and a room temperature \SI{10}{k\ohm} bias resistor.

We also have developed matching X-ray collimators, consisting of $(\SI{150}{\micro m})^2$ square apertures DRIE etched into Si wafers.
These are subsequently electroplated with \SI{10}{\micro m} of Au.
This thickness of gold is sufficient to stop 98\% of incoming \SI{8}{keV} photons, and so ensure the majority of transmitted photons are incident near the centre of the absorbers.
Features on the TES chip and collimator allow for the alignment of the collimator apertures with the TES absorbers, and for the collimator to rest on the TES chip without contacting the absorbers.
After clamping the collimator and TES chip assembly to a sample platform, gold wirebonds are added around the perimeter of both the collimator and TES chip to aid with thermalisation.
When mounted in the cryostat, the entire sample platform is enclosed in a high magnetic permeability shield to avoid trapping magnetic flux.

All fabrication steps are undertaken at the APS, Argonne's Centre for Nanoscale Materials (CNM), and the University of Chicago's Pritzker Nanofabrication Facility (PNF).
We have fabricated and tested a number of iterations of devices, with slight variations on TES design (e.g. number of Cu bars), absorber size, and membrane dimensions/perforations.
In this work, we focus on the characterisation and measurements of one particular 24-pixel die with a uniform TES design: three interdigitated Cu bars on the bilayer, and large $(\SI{750}{\micro m})^2$ absorbers.

\subsection{Readout}

The simultaneous readout of our TES arrays is done by microwave frequency multiplexing, specifically using the \umux{} microwave SQUID multiplexing chips designed and provided by NIST~\cite{Mates2008}.
Each multiplexing chip contains 33 SQUID amplifiers and \SI{300}{kHz} bandwidth superconducting microwave resonators with quality factors $Q > \num{1e4}$. %

The TES current for each pixel is inductively coupled to a SQUID on the multiplexing chip, which in turn is coupled to a microwave resonator.
Changes in the SQUID current cause a shift in the resonant frequency, and a change in the transmitted phase and amplitude of an on-resonance frequency tone.
The SQUID response to TES current is linearised by flux ramp modulation, in which a sawtooth-waveform current is also coupled into the SQUID array~\cite{Mates2012}, and the TES current is detected as a shift in the phase of the oscillations in the transmitted microwave signal.
The microwave signal is amplified by a low-noise HEMT amplifier at \SI{4}{K} and then by room temperature amplifiers.
In our lab setup, the typical noise of the readout system (SQUID, resonators, and amplifiers) is \SI{4}{\micro\Phi_0.Hz^{-1/2}}, where $\Phi_0$ is the magnetic flux quantum and corresponds to \SI{8.9}{\micro A} of TES current, below the TES noise level.

The room temperature electronics consist of a ROACH2 FPGA~\cite{ROACH2} and MUSIC ADC/DAC board~\cite{MUSIC}, along with a programmable local oscillator and appropriate IQ mixers, filters, and RF/IF amplifiers. The \umux{} readout firmware for the ROACH2 was also developed and provided by NIST~\cite{Gard2018}.
It is capable of generating and reading microwave tones for 128 channels simultaneously with a \SI{512}{MHz} bandwidth. The flux ramp demodulation for each channel is also performed within the ROACH2.
For the experiments reported here, we successfully read out the 33 channels of one \umux{} chip, sampling each channel (after demodulation) at \SI{62.5}{kHz} with 14-bit resolution.

\subsection{TES characterisation}

\begin{figure}[tbh]
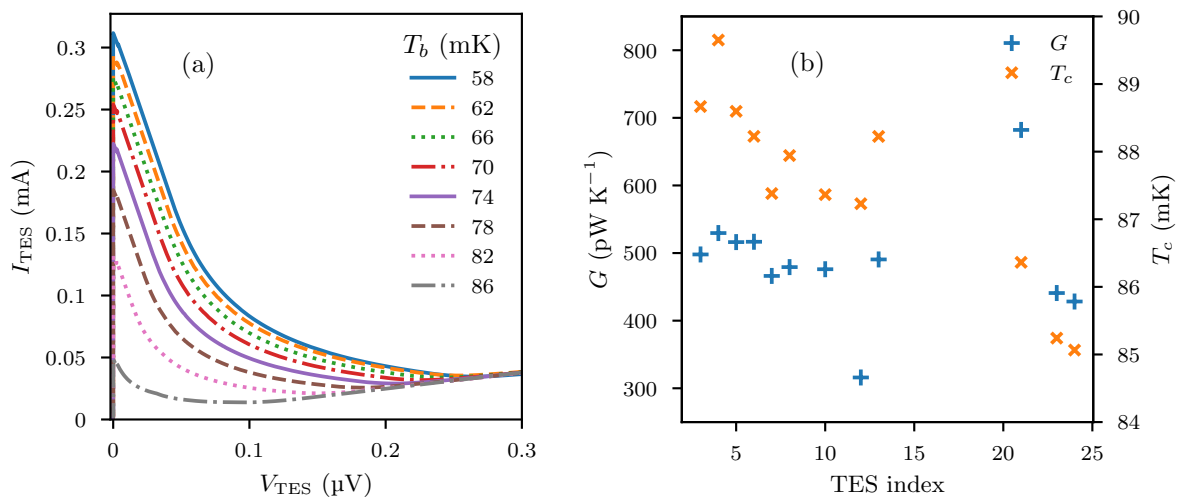

  \centering%
  \input{figures/iv_m452_die41.pgf}%
  \hspace{0.05in}%
  \input{figures/fitted_G_m452_die41.pgf}%
  \caption{(a)~Representative TES current-voltage measurements for one device, at a range of bath temperatures $T_b$. (b)~Fitted TES-bath thermal conductivities $G$ (blue $+$) and critical temperatures $T_c$ (orange $\times$) for 12 devices.\label{fig:g_fits}}
\end{figure}

TES current-voltage (IV) curves were measured at a range of temperatures in one of our ADR cryostats (base temperature \SI{55}{mK}); examples for one device are shown in \cref{fig:g_fits}(a).
During our tests, 12 of the 24 pixels on the die of interest showed superconducting transitions of the expected magnitude in the explored temperature range.
We believe the majority of the missing devices are due to wiring defects external to the TES chip, as they were also consistently absent when testing other dies.
The normal state resistances (mean $R_n = \SI{7.93}{m\ohm}$, standard deviation $\sigma = \SI{0.10}{m\ohm}$ of 12 pixels) and stray series resistances $R_p \sim \SI{7}{\micro\ohm}$ of each pixel were estimated from the IV slope at high voltage bias and low voltage bias (below the superconducting transition) respectively.
By fitting to the equation $P = k({T_c}^n - {T_b}^n)$ describing the power flow from TES to substrate, where we assume the TES is at its critical temperature $T_c$, $T_b$ is the bath temperature, and $P$ is the Joule power dissipated when the TES resistance $R\tu{TES} = 0.8 R_n$, we extracted the parameters $n$, $k$, and $T_c$ and estimated the thermal conductances from TES to bath $G = n k {T_c}^{n-1}$ for each pixel.
As illustrated in \cref{fig:g_fits}(b), the 12 pixels have a mean $T_c = \SI{87.5}{mK}$, $\sigma = \SI{1.4}{mK}$ and mean $G = \SI{487}{pW.K^{-1}}$, $\sigma = \SI{84}{pW.K^{-1}}$, with $G$ somewhat correlated with the varying $T_c$.
We are aware from separate tests of some non-uniformity in Mo thickness which causes $T_c$ to decrease as distance from the centre of the wafer increases. We are working to correct this by modifying our sputtering process~\cite{Patel2019sub}, which should reduce some of the variation between pixels.
In tests with other device designs, we find $G$ scales approximately linearly with the membrane perimeter.

X-ray pulse measurements at temperatures close to $T_c$, where electrothermal feedback is minimal, were used to estimate the characteristic thermal time constant $\tau = C/G$ and hence the heat capacity $C$ of each pixel.
This method has a large margin of error but suggests total heat capacities of around \SI{4}{pJ.K^{-1}} for this size of absorber.

\section{Energy spectra}

\subsection{Measurement setup}
For X-ray TES measurements, a tungsten filament X-ray tube source is mounted on the cryostat.
The source is oriented parallel to the cryostat wall, so that a sample placed in the beam can fluoresce towards the detectors at a 90$^\circ$ angle.
A circular area of the sample of about \SI{1}{cm} in diameter is illuminated by the X-ray beam.
The detector array is aligned with windows in the cryostat and magnetic shield (Be for the outer vacuum window, thin Al for the inner windows) to allow for photon transmission.
X-ray pulse measurements were taken with constant voltage bias for all pixels.
This implies the exact bias point ($R\tu{TES}/R_n$) varies slightly pixel-to-pixel.
A bias voltage of \SI{4}{V} corresponds to a bias of \SIrange{10}{22}{\%} for the 12 working pixels.
This range will reduce as the uniformity of $T_c$ and $G$ across our arrays improves.

To record pulses, the continuous digitized data stream from the ROACH2 was unwrapped (the \umux{} output is periodic over \SI{1}{\Phi_0}), then fed to a software derivative-based trigger and processed into pulse records of total length \SI{8}{ms} (500 samples), including \SI{2}{ms} of pre-trigger samples.
Pulse processing, including optimal filtering and corrections for known correlations were performed offline, primarily using the methods and MASS software framework developed at NIST~\cite{Fowler2016}.
The key steps in our offline pulse processing, done separately per pixel, are:
\begin{enumerate}
  \item Basic statistics (e.g. peak height, pretrigger mean, pretrigger RMS value, peak time, maximum post-trigger slope) are computed for each pulse record
  \item Pulse records which are significant outliers in pretrigger RMS value, peak time, and maximum post-trigger slope are discarded. These are typically records with pulse pileup (multiple pulses)
  \item An average pulse is calculated from pulses with heights close to the peak height median (the most prominent spectral line)
  \item The average pulse, along with the noise PSD, is used to create an optimal filter template
  \item This optimal filter template is convolved with each pulse record with 5 different time offsets, and the maximum of a fitted quadratic is used as the filtered pulse height. This accounts for sub-sample variations in the arrival time of the pulses
  \item The correlation between filtered pulse height and pretrigger mean is fitted and removed -- this accounts for slow changes in the gain of the detector, e.g. due to bath temperature variations
  \item Calibration points are found by matching peaks in the kernel density estimate (of the filtered, corrected pulse heights) with known prominent line energies
  \item Fits of each emission line, based on known intrinsic linewidths and a Gaussian detector response function, are performed to adjust the calibration points
  \item A power law function is fitted to these points and then used to convert all filtered, corrected pulse heights to calibrated energy
\end{enumerate}

\subsection{Energy resolution and linearity}

\begin{figure}[htb]
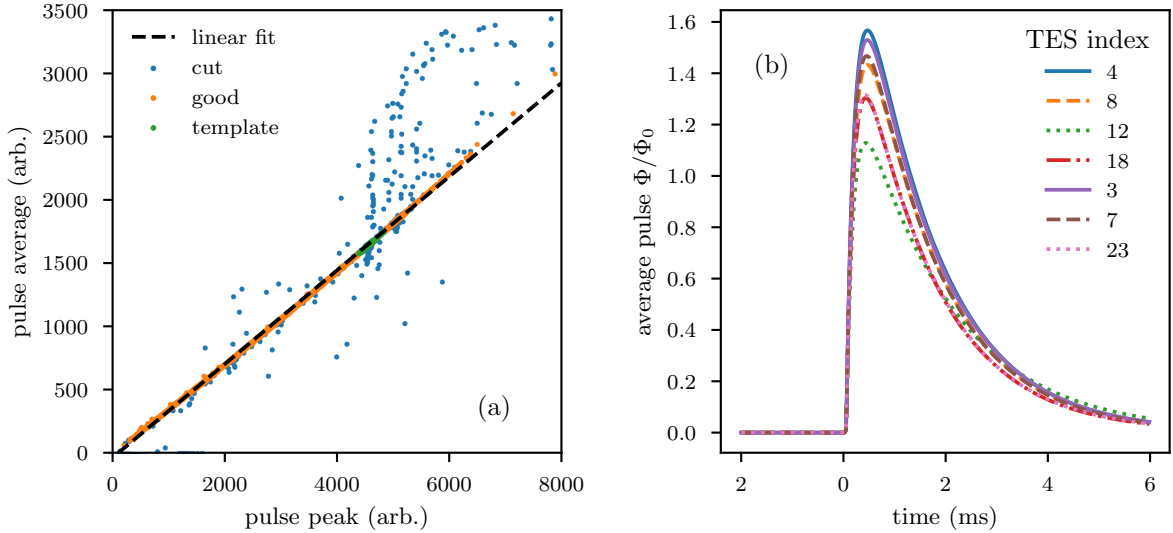

  \centering%
  \input{figures/mo_linearity_m452_die41.pgf}%
  \hspace{0.05in}%
  \input{figures/mo_average_pulses_m452_die41.pgf}%
  \caption{(a)~Average value of each pulse record against maximum value (for positive-going pulses) for one pixel. In blue are the pulses discarded as outliers, in orange the pulses kept to build the spectrum, and in green the Mo \kalpha{} pulses used to construct the optimal filter template.
  The black dashed line indicates a linear fit to the good pulses.
  (b)~The average pulse shapes for selected pixels in the array. These correspond to Mo \kalpha{} (\SI{17.5}{keV}) photons.
  Both figures are of fluorescence from a Mo foil sample, collected at \SI{4.0}{V} bias, \SI{62}{mK} bath temperature, and at an average of \SI{2}{counts/s} per pixel.\label{fig:mo_pulses}}
\end{figure}

Initial measurements were performed with metal foils (thickness \SIrange{0.5}{1}{mm}) of Ti (\kalpha{} $= \SI{4.5}{keV}$), Cu (\kalpha{} $= \SI{8}{keV}$), and Mo (\kalpha{} $= \SI{17.5}{keV}$).
After collecting pulses from the Mo foil, plotting the maximum value of each pulse record against the average value of each pulse record as in \cref{fig:mo_pulses}(a) shows a linear relationship until well past the Mo \kalpha{} emission line, i.e. the pulse shape is constant over this range.
Based on this, we estimate these pixels have a saturation energy $E\tu{sat} > \SI{20}{keV}$.
The total height and slew rate of the pulses, illustrated in \cref{fig:mo_pulses}(b), are well within the capabilities of our readout system when sampling at \SI{62.5}{kHz}.

\begin{figure}[htb]
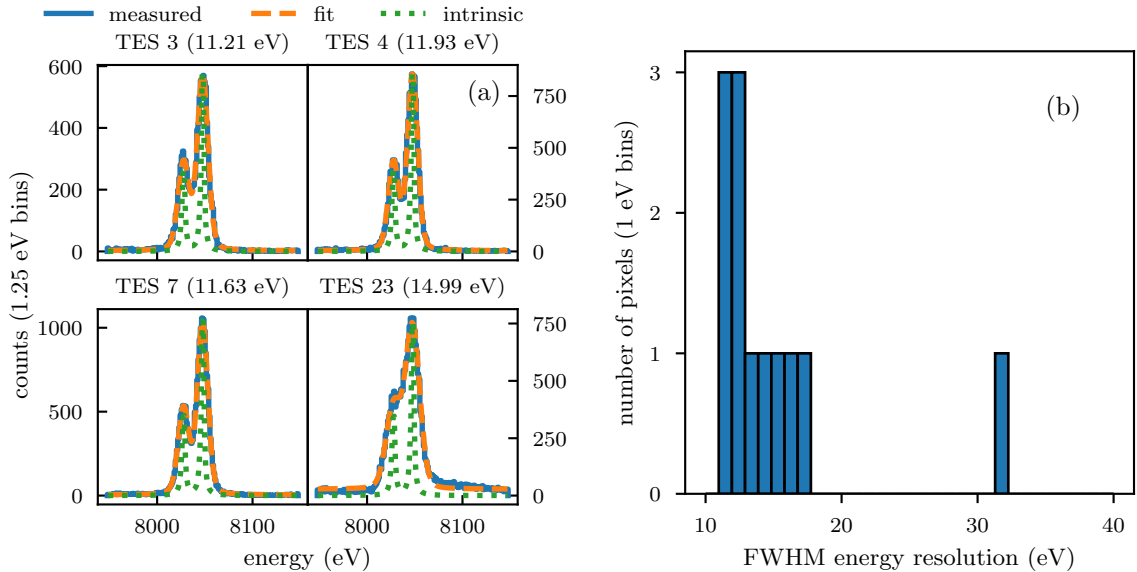

  \centering%
  \input{figures/cu_per_pixel_fits_m452_die41.pgf}%
  \hspace{0.05in}%
  \input{figures/cu_resolution_histogram_m452_die41.pgf}%
  \caption{(a)~Calibrated pulse energy histograms (solid blue) and fitted functions (dashed orange) accounting for the intrinsic linewidth of the Cu \kalpha{} line (dotted green) and Gaussian broadening due to the detector, for selected pixels.
  (b)~Histogram of FWHM energy resolution for 12 pixels on the tested die, from Cu \kalpha{} (\SI{8}{keV}) emission line fits.
  Data collected at \SI{4.25}{V} bias, \SI{62}{mK} bath temperature and $\sim \SI{2}{counts/s}$ per pixel.\label{fig:cu_fits}}
\end{figure}

Fitting the known line shape of the Cu \kalpha{} complex~\cite{Holzer1997}, convolved with a Gaussian of varying width, to the measured spectra gives the per-pixel FWHM energy resolutions shown in \cref{fig:cu_fits}.
In all cases the fitted resolution is close (most within \SI{0.5}{eV}) to the optimal filter predicted resolution based on the average pulse shape and noise PSD (under X-ray illumination, with pulses removed) for that pixel.
All measured pixels in the array have a resolution of \SIrange{11}{17}{eV}, with 6 of 12 below \SI{13}{eV}, except one which has unexpectedly high noise and consequently poor energy resolution ($>\SI{30}{eV}$).

There was a factor of three variation in input count rates between pixels, probably due to the X-ray window not being exactly centred over the array.
However, the energy resolution was not strongly correlated with the individual pixel count rate.
Instead, energy resolution noticeably degraded as the total (for the whole array) incident photon rate was increased; by increasing the total detected photon flux from \SI{25}{counts/s} to \SI{85}{counts/s} for the whole array, the typical energy resolution worsened from \SI{12}{eV} to \SI{15}{eV}.
We found that improving the thermal contact of the chip to the sample platform, e.g. by adding gold wirebonds, reduced this effect, as did increasing the thickness of gold on the collimator.
This suggests that stray X-rays absorbed in the bulk of the chip are causing sufficient temperature fluctuations to affect our measured energy resolution.
Currently, of the \SI{12}{eV} typical energy resolution, we attribute around \SI{3}{eV} to this flux dependent effect, based on the differences in the measured noise PSDs with and without X-ray photons.
We are investigating methods to reduce this effect in the next iteration of our devices.
We also find that the average pulse height is very well correlated with the per pixel bias point, and weakly correlated with measured energy resolution.
Therefore, the variation in measured energy resolution is at least partly due to the variation in $G$ and $T_c$ between pixels. %

\subsection{Multi-element samples and comparison with SDDs}

\begin{figure}[htb]
  \centering%
  \input{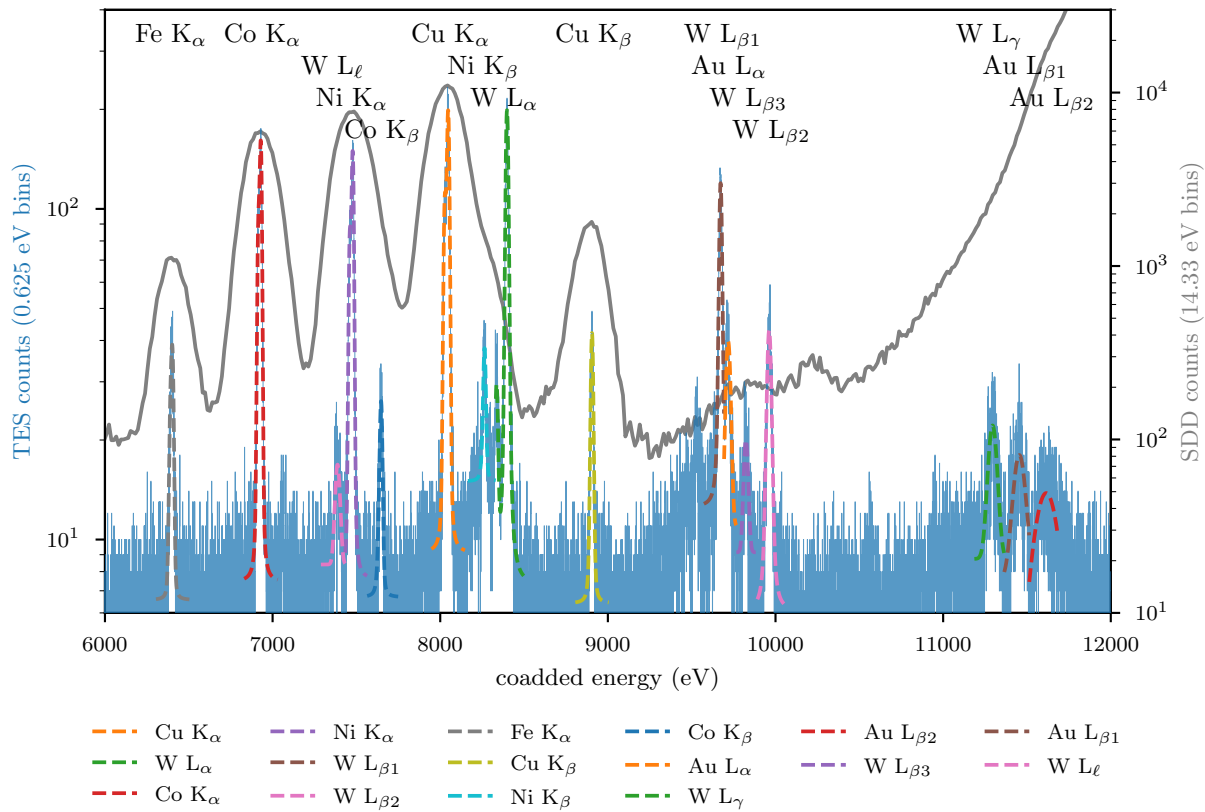}%
  \caption{Co-added (12 pixels) calibrated fluorescence spectrum from a Cu-Ni-Co thin film, measured by our TES array (blue, left axis) and a Vortex SDD at a beamline (grey, right axis).
  Dashed lines indicate approximate fits to emission lines visible in the TES-measured spectrum.
  TES data collected at \SI{4.25}{V} bias, \SI{62}{mK} bath temperature and $\sim \SI{2}{counts/s}$ per pixel.\label{fig:cunico}}
\end{figure}

A trilayer thin film of Cu, Ni, and Co in a $1:1:1$ mass ratio on quartz substrate was chosen as a simple demonstration of the capabilities of a TES-based spectrometer.
This sample, when measured with a Hitachi Vortex single-pixel SDD at the 1-BM beamline of the APS, shows (\cref{fig:cunico}) a spectrum with the three \kalpha{} peaks clear, but the \kalpha[\beta] lines of Co and Ni peaks completely hidden.
Also visible in the SDD-measured spectrum is elastic scattering from the monochromatic synchrotron beam at \SI{12}{keV} and some background Fe.
Measuring with our lab X-ray tube source and TES array at an average of $\SI{2}{counts/s}$ per pixel, we are able to clearly resolve the three \kalpha[\beta] peaks, separated from each other and from the emission lines from the X-ray tube W anode.
Fitting the Cu \kalpha{} line in the co-added spectrum (combining data from all 12 pixels) yields $\Delta E\tu{FWHM} = \SI{15}{eV}$.
In a partial fluorescence yield X-ray absorption measurement of this sample, one would not be able to correctly measure the Ni absorption edge via the \kalpha{} line because of the overlapping Co \kalpha[\beta] line using an SDD, but would be able to with a TES array.

\begin{figure}[htb]
  \centering%
  \input{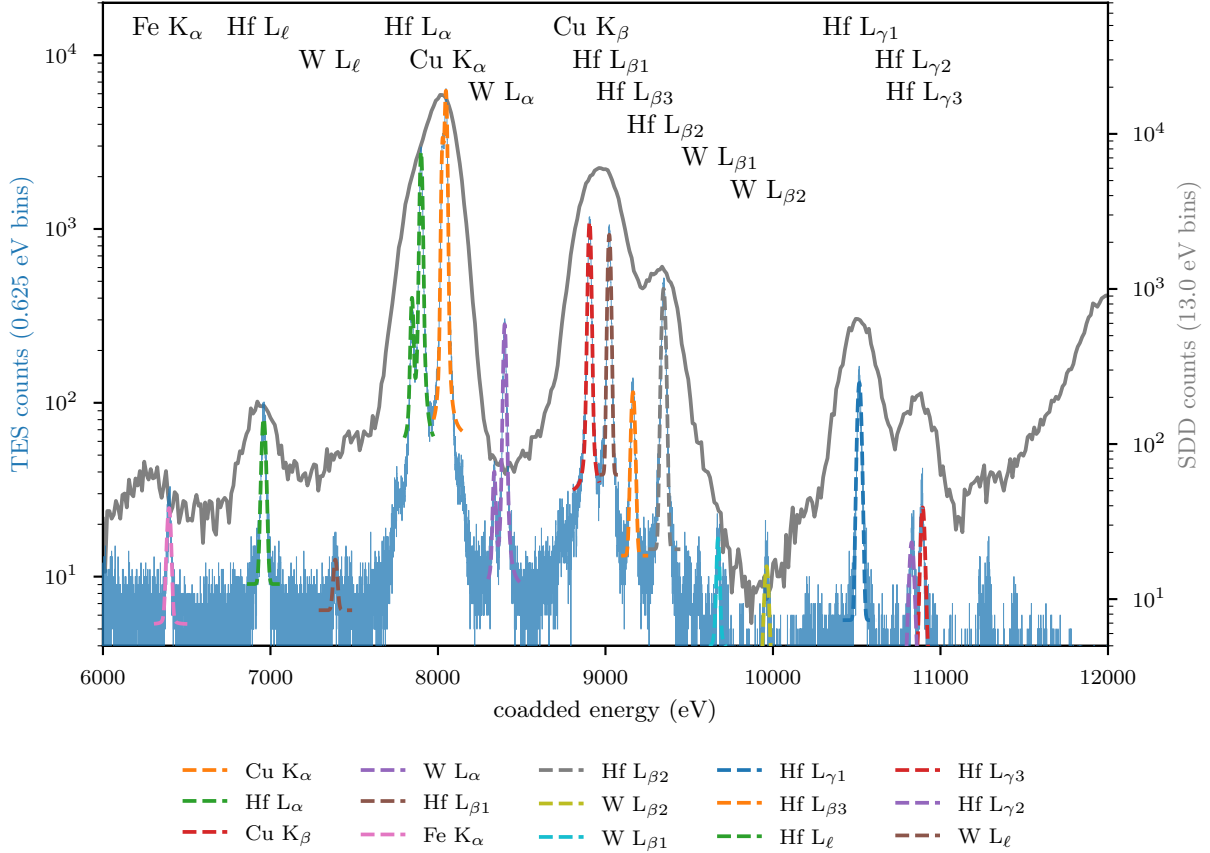}%
  \caption{Co-added (12 pixels) calibrated fluorescence spectrum from the simultaneous illumination of Cu and Hf foils, measured by our TES array (blue, left axis), and a Vortex SDD at a beamline (grey, right axis). Dashed lines indicate approximate fits to emission lines visible in the TES-measured spectrum.
  TES data collected at \SI{4.25}{V} bias, \SI{62}{mK} bath temperature and $\sim \SI{2}{counts/s}$ per pixel.\label{fig:cuhf}}
\end{figure}

In another experiment, Cu and Hf foils were placed next to each other and simultaneously illuminated. The Hf \lalpha{} line lies on the low-energy shoulder of the Cu \kalpha{} peak and is poorly resolved by an SDD.
The TES array is able to clearly resolve the two, and also separate out the Cu \kalpha[\beta] and several Hf \lalpha[\beta] lines, again measuring at an average of $\SI{2}{counts/s}$ per pixel.
The 12 pixel co-added spectrum is shown in \cref{fig:cuhf} in comparison with the spectrum measured with a Hitachi Vortex SDD.
Once again the \SI{12}{keV} beam is only visible in the SDD spectrum, while the W emission lines are only visible in the TES spectrum.
As compared to the thin film, the thicker foils produce less elastic scatter, and so a lower background in both measurements of the spectrum.
Spectra of this nature will arise in fluorescence mapping of integrated circuits. An SDD would likely be unable to identify the prescence of Hf amidst a strong Cu background.

\section{Conclusions}
Prototype 24-pixel Transition Edge Sensor arrays for hard X-ray beamline science have been fabricated and tested at the Advanced Photon Source.
Here we measured devices with saturation energies $>\SI{20}{keV}$, and calculated the energy resolutions of most pixels to be \SIrange{11}{14}{eV} FWHM at \SI{8}{keV}.
This energy range is of importance to the APS, and relatively unexplored with TESs at synchrotron beamlines.
The measured energy resolution is already sufficient to enable new X-ray spectroscopy science, as demonstrated by measurements of the fluorescence spectra of a Cu-Ni-Co thin film and foils of Cu and Hf.
These are representative of real samples such as integrated circuits and organometallic complexes in biological samples.
As compared to the same spectra measured with a silicon drift detector at an APS beamline, the TES-measured spectra show clearly resolved prominent emission lines and reveal several weaker ones.
Work is ongoing to increase the pixel count of the array, improve uniformity, and reduce the effect of photon flux on energy resolution.
We also intend to collect emission spectra from more samples, such as integrated circuit dies fabricated with different process nodes, to further explore the capabilities of a TES array at a beamline.
These measurements provide a simple demonstration that when complete, this first-generation instrument will be valuable for X-ray fluorescence and absorption spectroscopy at the APS.

\ack
We thank D. T. Becker, D. A. Bennett, J. A. B. Mates, and K. M. Morgan for useful discussions, and the Quantum Sensors Group at NIST (Boulder, CO, USA) for providing and helping implement the \umux{} readout.
We also thank H. Jian and X. Zhang for providing the Cu-Ni-Co thin film sample.
The silicon drift detector measurements were performed at the 1-BM beamline of the APS with the assistance of M. Wojcik.
D. Czaplewski, R. Divan, S. Miller, and O. Makarova provided advice on fabrication.
This research is funded by Argonne National Laboratory LDRD proposal 2018-002-N0: Development of a Hard X-ray Spectrometer Based on Transition Edge Sensors for Advanced Spectroscopy; 
was supported by the Accelerator and Detector R\&D program in Basic Energy Sciences' Scientific User Facilities (SUF) Division at the Department of Energy;
used resources of the Advanced Photon Source and Center for Nanoscale Materials, U.S. Department of Energy Office of Science User Facilities operated for the DOE Office of Science by the Argonne National Laboratory under Contract No. DE-AC02-06CH11357;
and made use of the Pritzker Nanofabrication Facility of the Institute for Molecular Engineering at the University of Chicago, which receives support from Soft and Hybrid Nanotechnology Experimental (SHyNE) Resource (NSF ECCS-1542205), a node of the National Science Foundation’s National Nanotechnology Coordinated Infrastructure.

\bibliography{proceedings}

\providecommand{\newblock}{}
\begin{thebibliography}{10}
\expandafter\ifx\csname url\endcsname\relax
  \def\url#1{{\tt #1}}\fi
\expandafter\ifx\csname urlprefix\endcsname\relax\def\urlprefix{URL }\fi
\providecommand{\eprint}[2][]{\url{#2}}

\bibitem{Doriese2017}
Doriese W~B {\em et~al.\/} 2017 {\em Review of Scientific Instruments\/} {\bf
  88} 053108

\bibitem{Bennett2012}
Bennett D~A {\em et~al.\/} 2012 {\em Review of Scientific Instruments\/} {\bf
  83} 093113

\bibitem{Hatakeyama2014}
Hatakeyama S, Ohno M, Takahashi H, Damayanthi R~M, Otani C, Yasumune T, Ohnishi
  T, Takasaki K and Koyama S 2014 {\em Journal of Low Temperature Physics\/}
  {\bf 176} 560--565

\bibitem{Mates2017}
Mates J~A~B {\em et~al.\/} 2017 {\em Applied Physics Letters\/} {\bf 111}
  062601

\bibitem{Choi2011}
Choi J~H, Mao Y and Chang J~P 2011 {\em Materials Science and Engineering R:
  Reports\/} {\bf 72} 97--136

\bibitem{Davis2015}
Davis K~M and Pushkar Y~N 2015 {\em Journal of Physical Chemistry B\/} {\bf
  119} 3492--3498

\bibitem{Vortex}
{Hitachi High-Technologies Science America} 2019 {Vortex X-ray Detector}
  \urlprefix\url{https://www.hitachi-hightech.com/hhs-us/product_detail/?pn=ana-vortex}

\bibitem{Ullom2004}
Ullom J~N, Doriese W~B, Hilton G~C, Beall J~A, Deiker S, Duncan W~D, Ferreira
  L, Irwin K~D, Reintsema C~D and Vale L~R 2004 {\em Applied Physics Letters\/}
  {\bf 84} 4206--4208

\bibitem{Yan2017a}
Yan D {\em et~al.\/} 2017 {\em Applied Physics Letters\/} {\bf 111} 192602

\bibitem{Yan2018}
Yan D {\em et~al.\/} 2018 {\em Journal of Low Temperature Physics\/} {\bf 193}
  225--230

\bibitem{Patel2019sub}
Patel U, Divan R, Gades L, Guruswamy T, Yan D, Quaranta O and Miceli A 2019
  {\em Journal of Low Temperature Physics\/}  (submitted) (\textit{Preprint}
  \eprint{1908.10239})

\bibitem{Mates2008}
Mates J~A~B, Hilton G~C, Irwin K~D, Vale L~R and Lehnert K~W 2008 {\em Applied
  Physics Letters\/} {\bf 92} 023514

\bibitem{Mates2012}
Mates J~A~B, Irwin K~D, Vale L~R, Hilton G~C, Gao J and Lehnert K~W 2012 {\em
  Journal of Low Temperature Physics\/} {\bf 167} 707--712

\bibitem{ROACH2}
{Collaboration for Astronomy Signal Processing and Electronics Research
  (CASPER)} 2013 {ROACH-2 Revision 2}
  \urlprefix\url{https://casper.ssl.berkeley.edu/wiki/ROACH-2_Revision_2}

\bibitem{MUSIC}
{Collaboration for Astronomy Signal Processing and Electronics Research
  (CASPER)} 2015 {MUSIC Readout}
  \urlprefix\url{https://casper.ssl.berkeley.edu/wiki/MUSIC_Readout_(Kinetic_Inductance_Detector_(KIDs))}

\bibitem{Gard2018}
Gard J~D, Becker D~T, Bennett D~A, Fowler J~W, Hilton G~C, Mates J~A, Reintsema
  C~D, Schmidt D~R, Swetz D~S and Ullom J~N 2018 {\em Journal of Low
  Temperature Physics\/} {\bf 193} 485--497

\bibitem{Fowler2016}
Fowler J~W, Alpert B~K, Doriese W~B, Joe Y~I, O'Neil G~C, Ullom J~N and Swetz
  D~S 2016 {\em Journal of Low Temperature Physics\/} {\bf 184} 374--381

\bibitem{Holzer1997}
H{\"{o}}lzer G, Fritsch M, Deutsch M, H{\"{a}}rtwig J and F{\"{o}}rster E 1997
  {\em Physical Review A\/} {\bf 56} 4554--4568

\end{thebibliography}

\end{document}